\documentclass[doublecol]{epl2} 

\usepackage{amsfonts,amsmath,amssymb}
\usepackage{graphics,psfrag}
\usepackage{graphicx,psfrag}
\usepackage{color,soul}


\title{Exact results for the extreme Thouless effect in a model of network dynamics}
\shorttitle{Extreme Thouless effect} 

\author{ R.K.P. Zia\inst{1} \and Weibin Zhang\inst{2,3} \and Mohammadmehdi Ezzatabadipour\inst{2,3} \and Kevin E. Bassler\inst{2,3,4}}

\shortauthor{R.K.P. Zia, \etal}

\institute{
  \inst{1}
Center for Soft Matter and Biological Physics, 
Department of Physics, Virginia Polytechnic Institute and State University, Blacksburg, VA 24061, USA\\
  \inst{2}
Department of Physics, University of Houston, Houston, Texas 77204, USA\\
  \inst{3}
Texas Center for Superconductivity, University of Houston, Houston, Texas 77204, USA\\
  \inst{4} 
Department of Mathematics, University of Houston, Houston, Texas 77204, USA}

\abstract{
If a system undergoing phase transitions exhibits some characteristics of
both first and second order, it is said to be of `mixed order' or to display
the Thouless effect. Such a transition is present in a simple model of a dynamic social network, in which $N_{I/E}$ extreme introverts/extroverts
always cut/add random links. In particular, simulations
showed that $\left\langle f\right\rangle $, the average fraction of cross-links between the two groups
(which serves as an `order parameter' here), 
jumps dramatically when $\Delta \equiv N_{I}-N_{E}$ crosses the `critical
point' $\Delta _{c}=0$, as in typical first order transitions. 
Yet, at criticality, there is no phase co-existence, 
but the fluctuations of $f$ are much larger than in typical second order transitions. 
Indeed, it was conjectured that, in the thermodynamic limit, both
the jump and the fluctuations become maximal, so that the system is said to
display an `extreme Thouless effect.' While earlier theories are partially
successful, we provide a mean-field like approach that accounts for all
known simulation data and validates the conjecture. Moreover, for the
critical system $N_{I}=N_{E}=L$, an analytic expression for the mesa-like
stationary distribution, $P\left( f\right) $, shows that it is essentially
flat in a range $\left[ f_{0},1-f_{0}\right] $, with $f_0 \ll 1$. Numerical evaluations of $f_{0}$
provides excellent agreement with simulation data for $L\lesssim 2000$. 
For large $L$, we find $f_{0}\rightarrow \sqrt{\left(  \ln L^2 \right) /L}$ , 
though this behavior begins to set in only for $L>10^{100}$. 
For accessible values of $L$, we provide a transcendental equation for an approximate $f_{0}$ which is
better than $\sim$1\% down to $L=100$. We conjecture how this approach might be used to attack
other systems displaying an extreme Thouless effect.
}

\begin{document}

\maketitle

\section{Introduction}

Phase transitions are dramatic occurrences of collective behavior in systems
with large numbers of degrees of freedom ($\mathcal{N}$). They are ubiquitous,
while nearly all of life on earth depends on their existence (e.g., the
ice-water-vapor transitions). Based on the works of Boltzmann, Gibbs, and
Ehrenfest, textbook treatments focus mostly on first and second order
transitions, emphasizing on the different characteristics. Typically, an order
parameter (OP) is defined so that it is finite as $\mathcal{N}\rightarrow
\infty$ (the thermodynamic limit) and its dependence on the control parameters
(CP) displays different behaviors in the various phases. As the CPs are
varied across these transitions
(e.g., at the water-vapor transition across 100$^{\circ}$C under 1
atm pressure or the Curie point for ferromagnetic systems), 
the OP or its derivative suffers a discontinuity. The Lenz-Ising
system \cite{Ising1925}, with external field and temperature as CPs, is
arguably the simplest theoretical model which is known to display both of
these transitions \cite{Onsanger44,McCoyWu73}.

Many characteristics of these transitions are commonly accepted. Though the OP
is singular (discontinuous) at a first order transition, its derivatives
remain finite (on either side of the transition). Since these derivatives are
associated with the fluctuations of the OP in the system, the implication is
that `normal' Gaussian fluctuations (as $\mathcal{N}\rightarrow\infty$)
prevail, along with the notion of finite correlation lengths. At the
transition itself, the system may display phase co-existence (e.g., water and
steam at 100$^{\circ}$C), if the system is constrained so that the OP is
forced to be a value within the discontinuity. Deep inside each phase, the fluctuations and the correlation lengths are \textit{finite},
taking on the values on either side of the transition. By contrast, as a CP crosses
a second order transition, the OP remains continuous, but its
derivative displays a discontinuity. Often, this discontinuity is infinite,
diverging with some non-rational exponent (critical exponent). In other words,
the fluctuations of the OP and the correlation length typically become
`anomalously divergent.' Finite size scaling is a well-established method
\cite{vladimir1990finite} that displays clearly how the OP behaves as a function of
both $\mathcal{N}$ and the CPs. 

Studying one-dimensional Ising models with long-range interactions,
Thouless\cite{Thouless69} found that some systems do not follow such `standard
behavior.' In particular, the OP may jump discontinuously \textit{and} large 
fluctuations exist at criticality. Many systems displaying such
`mixed-order transitions' have been found
\cite{ poland_phase_1966,fisher_effect_1966, yuval1970exact, aizenman1988discontinuity,blossey1995diverging, kafri_why_2000, toninelli2006jamming, schwarz2006onset, bizhani2012discontinuous, whitehouse2014maintenance,  FFK16, FronczakFronczak16, choi2017mixed,juhasz2017mixed, alert2017mixed}. Most recently, the term `extreme
Thouless effect' has been coined\cite{BM14a,BM14b} for systems in which both
the discontinuity and the fluctuations are maximal (e.g., the magnetization in
an Ising-like model jumping from $-1$ to $+1$). Such behavior has been
observed also in a model of social dynamics in which two subgroups of
individuals always cut/add links (`extreme introverts and extroverts' or
$XIE$) \cite{LSZepl2012,BLSZpre15}. For this simple model, the only 
CPs are $N_{I/E}$, the numbers of introverts/extroverts, with
$L \equiv \left( N_{I} + N_{E} \right) / 2$ and $\Delta \equiv N_{E} - N_{I}$ as natural alternatives.
Meanwhile, the fraction of cross-links, $f$, plays the role of an OP.
In simulations with $L=100$, a dramatic jump in $f$ was observed when $\Delta$ crosses 
the `critical' value $\Delta _{c} = 0$, giving the impression of a first order transition. Yet, at criticality, the fluctuations of $f$ are non-Gaussian and large 
(comparable to the jump in magnitude), more typical of second order transitions. 
The extreme Thouless effect is based on extrapolations with data on systems with $L\lesssim2000$, but with little understanding of the $L\rightarrow\infty$ limit.
Theoretical arguments have been put forth\cite{BDZ15}, suggesting 
that the jump approaches unity with $O\left(  \sqrt{1/L}\right)  $ corrections.
This letter is devoted to a fresh analytic approach, providing 
exact results which agree well with all simulation data. In particular, the approach is
found to be quite subtle, following more closely the solution of 
a transcendental equation for $L$'s accessible to our computers and converging onto the asymptote 
$\sqrt{\left(  \ln L^2 \right) /L}$ at hopelessly large $L$'s. The next section is a brief summary of
the $XIE$ model, followed by some details of the novel analysis. Given these 
insights on $XIE$, we speculate in a final section on possible avenues for
research in other systems that display an extreme Thouless effect.

\section{A simple model ($XIE$) of dynamic networks and the extreme Thouless
effect}

In a typical social network, links between individuals are dynamic, as new
ones are created while others are cut. At any time, the topology is completely
specified by the adjacency matrix, $\mathbb{A}$, an element of which,
$a_{ij}$, is unity or zero depending on the presence or absence of a link
between individuals $i$ and $j$ (and so, $a_{ij} = a_{ji}$). Thus, $\mathbb{A}\left(  t\right)  $
represents an evolution trajectory of the network and resembles an Ising model
on a square lattice (with spins $2a_{ij}-1$). Now, it is natural for some
individuals to prefer large numbers of contacts and others, few friends. To
model such behavior, we introduced the notion of preferred degrees -- the
number of links with which an individual is most content. For example,
introverts (I) prefer few friends while extroverts (E) prefer many
contacts. While a wide distribution of such preferences can be found in a real
society, we focus on simple models in order to gain quantitative insight into
their effects on the network. In the simplest case, all individuals prefer the
same degree, $\kappa$. The stochastic evolution of our model involves
choosing a random individual and, if it has more than $\kappa$ links, it cuts
one of its existing links. If it has $\kappa$ or fewer links,
it adds a link to a random individual not already connected to it. Despite the
apparently random nature of this dynamic network, the system settles into a 
(non-equilibrium) stationary state, with the probability for finding 
$\mathbb{A}$,   $\mathcal{P}^{\ast}\left(  \mathbb{A}\right)$, which 
differs considerably from the Erd\H{o}s-R\'{e}nyi
distribution \cite{LJSZ2013,LSZ2014}. Apart from minor fluctuations, everyone is
more or less content with the `status quo.' Exploring the next simplest
system, we considered just two subgroups (e.g., I's and E's) with $\kappa_{1,2}$ 
\cite{LSZ2014} and discovered a number of surprising properties, including
anomalously large fluctuations in $X$, the total number of links between the
subgroups. As expected, we find the phenomenon of `frustration,' where some 
individuals are not content with the `status quo.'

A remarkable simplification of such two-subgroup networks emerges when we set
the $\kappa$'s at extreme values: zero and infinity.  
Coined the $XIE$ model, an I/E always attempts to cut/add links, so that the 
stationary state has no I-I links and all E-E links. 
Labeling $a_{ij}$ so that all indices for the I's are smaller than those for the E's, 
we see that $\mathbb{A}$ is a 2x2 block matrix with frozen I-I and E-E blocks. Only the I-E block remains dynamic,
representing the incident matrix of a bipartite graph: $\mathbb{N}$. This
$\mathbb{N}$ now plays the role of a rectangular ($N_{I}\times N_{E}$) Ising-type model. 
Meanwhile, $X$ is just the sum over all its elements and so, the
average fraction of cross-links, $\left\langle f\right\rangle \equiv
\left\langle X\right\rangle /N_{I}N_{E}$, plays the role of 
magnetization~\footnote{ Unlike the Ising model, the CPs here are the sizes of 
the system ($N_{I,E}$), though it is possible to introduce new CPs that 
correspond to the magnetic field and temperature.}. 
Remarkably, detailed balance is restored and the Boltzmann-like stationary distribution, 
$\mathcal{P}^{\ast}\left( \mathbb{N}\right)  $, was found analytically \cite{LSZepl2012}. 
Not surprisingly, the `Hamiltonian' 
$\mathcal{H}\equiv-\ln\mathcal{P}^{\ast}$ involves long-range and multi-spin 
interactions, so that it is a gargantuan challenge
to find analytically the `partition function', averages of observable
quantities, or the full distribution $P\left(  f\right)  \equiv\sum_{\left\{
\mathbb{N}\right\}  }\delta\left(  f-X/N_{I}N_{E}\right)  \mathcal{P}^{\ast
}\left(  \mathbb{N}\right)  $. On the other hand, it is straightforward to
perform Monte Carlo simulations. Employing systems with $L= 100$,  
$\left\langle f\right\rangle$ was discovered to jump from $0.14$ to $0.86$ 
when $\Delta$ changes from $-2$ to $+2$ (i.e.,
from 99 to 101 extroverts)\cite{LSZepl2012}! 
At $\Delta_{c}$, $\left\langle f\right\rangle =0.5$ by symmetry, 
but $f\left(  t\right)  $ resembles that of an unbiased 
random walk, confined between `soft walls' at approximately $0.21$ and $0.79$. 
In other words, the distribution $P\left(  f\right)  $ resembles a wide mesa, so 
that the variance of $f$ is $O\left(  1\right)  $, instead of the typical
$O\left( 1/L \right)  $ in a Landau theory~\footnote{For example, the Landau free energy for the Ising model is 
$\mathcal{N} f(m)$, with $f= \tau m^2 + u m^4$. 
As a result, far from criticality, $\tau > 0$ so that $m^2$ and the variance scale as $1/\mathcal{N}$. 
But at criticality, these scale as $ 1/ \sqrt{ \mathcal{N} } $. 
In two dimensions, $\mathcal{N} = L^2 $, giving us the $1/L$ in the text.}.
Such a combination (discontinuous OP and anomalously large fluctuations) is the
signature of a Thouless effect. The simplest mean-field analysis consists of
replacing every matrix element in $\mathcal{H}\left(  \mathbb{N}\right)  $ by
its average $\mu\in\left[  0,1\right]  $ and obtaining a Landau-like `free
energy' $\mathcal{F}\left(  \mu;N_{E},N_{I}\right)  $, so that $\left\langle f\right\rangle $ is identified by the minimum of $\mathcal{F}$.
Now, $\mathcal{F}=\mu \ln\left(  N_{I}/N_{E}\right)  $ at the lowest order in $L$, 
so that its minimum is $0$ (or $1$) for $\Delta <0$ (or $>0$). 
The conclusion is that this model displays an `extreme' Thouless effect \cite{LSZepl2012}. 
However, adding the next order (so that `soft walls' are present) did not produce quantitatively convincing results. 
In particular, as $L\rightarrow\infty$, $\left\langle f\right\rangle $ jumps from
$0.19$ to $0.81$ when $\Delta$ goes from $-2$ to $+2$. In other words, the 
`walls' approach \textit{constants, away} from the extremes. 

In subsequent studies\cite{LGBSZ14,BLSZpre15,BDZ15}, progress in simulations and theory indicate otherwise: 
The jumps in $f$ trend towards $0$ and $1$ for larger systems, while the `walls' 
in a \textit{critical} system are found to
approach these extremes. Relying on data with various $\Delta$ and $L$'s up to $1600$, 
rough scaling plots of $\,\left\langle f\right\rangle $ hint at
anomalous behavior, though it was difficult to arrive at reliable critical exponents. 
On the theoretical front, a self consistent mean-field (SCMF) theory was developed, 
focusing on the degree distributions of the two subgroups: $\rho_{I,E}\left(  k\right)$. 
The agreement with data were quite good, for all \textit{non-critical} systems\cite{BLSZpre15}. 
For the \textit{critical} system however, though qualitatively correct, 
the predictions are far from ideal. (See Fig. \ref{criticalDD} below.) 
Nevertheless, it was argued\cite{BDZ15} that the `walls' in this system should approach 
the extremes as $O\left(  \sqrt{1/L}\right)  $. 

\section{Theoretical studies, exact results and comparisons with simulations}

Here, we present a fresh perspective on the $XIE$ model, exploiting the ideas
of the SCMF theory\cite{BLSZpre15} in a different context. Instead of keeping only $N_{I,E}$ fixed 
and letting a self-consistency condition to determine $X$, we consider `cross sections' 
of the critical system ($N_{I}=N_{E}=L$) with fixed $X$, or $f=X/L^{2}$.
Such systems are similar to the lattice gas version of the Ising model \cite{YangLee52}  
in which the total magnetization is constrained. 
Clearly, simulations can be easily carried out for such ensembles. 
We show next how our new perspective leads to significant progress on the theoretical front.

Focusing on the steady state and to be specific, we consider an I with $k$ links 
and degree distribution in a fixed $X$ ensemble: $\rho_{I}\left(  k;X\right)  $. 
If chosen (with probability $1/2L$), the I will cut one of its links, \textit{unless} $k=0$. 
Thus, $\rho_{I}\left(  0;X\right) $ will play a crucial role. 
An I with $k-1$ links will gain a link if an E (not already connected to it) is chosen to act 
\textit{and} chooses to add a link to our particular I. The probability for 
these choices are, respectively, $\left(  L-k+1\right)  /\left(  2L\right)  $ and  
$1/\left( L-\ell\right) $, where $\ell$ is the number of I's already linked to this E. In
general, $\ell$ is a stochastic variable, but in the spirit of mean field
theory, we replace it by $\left\langle \ell\right\rangle =X/L$. To emphasize, 
this is just a constant in a fixed $X$ ensemble. 
Thus, we find $\rho _{I}$ explicitly by balancing gain and loss:%
\begin{equation}
\rho _{I}\left( k;X\right) =\frac{L!\left( L-X/L\right) ^{-k}}{Z\left(
X\right) \left( L-k\right) !}  \label{rho(k;X)}
\end{equation}%
where%
\begin{equation}
Z\left( X\right) =\frac{L!}{\left[ L\left( 1-f\right) \right] ^{L}}%
\sum_{q=0}^{L}\frac{\left[ L\left( 1-f\right) \right] ^{q}}{q!}
\label{Z(X)}
\end{equation}%
From the partial sum of an exponential series, it is clear that the limits of $L\rightarrow
\infty $ and $f\rightarrow 0$ do not commute. Provided $f$ is bounded from $0
$, we can prove that the sum approaches 
$ \exp \left[ L\left( 1-f\right) \right] $ as $L\rightarrow \infty $.

Armed with $\rho _{I}\left( 0;X\right) =1/Z\left( X\right) $, we return to the original critical system, in which $X$
wanders over most of its allowed values. As noted previously \cite{LSZepl2012},
that $X$ essentially performs an unbiased random walk (between `soft walls') can be understood as
follows. When $X$ is not close to $0$ or $L^2$, the I's have many links to cut
and the E's can add links to many unconnected I's. Thus, choosing any
individual (with equal probability) will result in $X$ changing by unity, so that 
$P\left( X\right) = P\left( X-1\right) $.

\begin{figure}[h!]
\includegraphics[width=\linewidth]{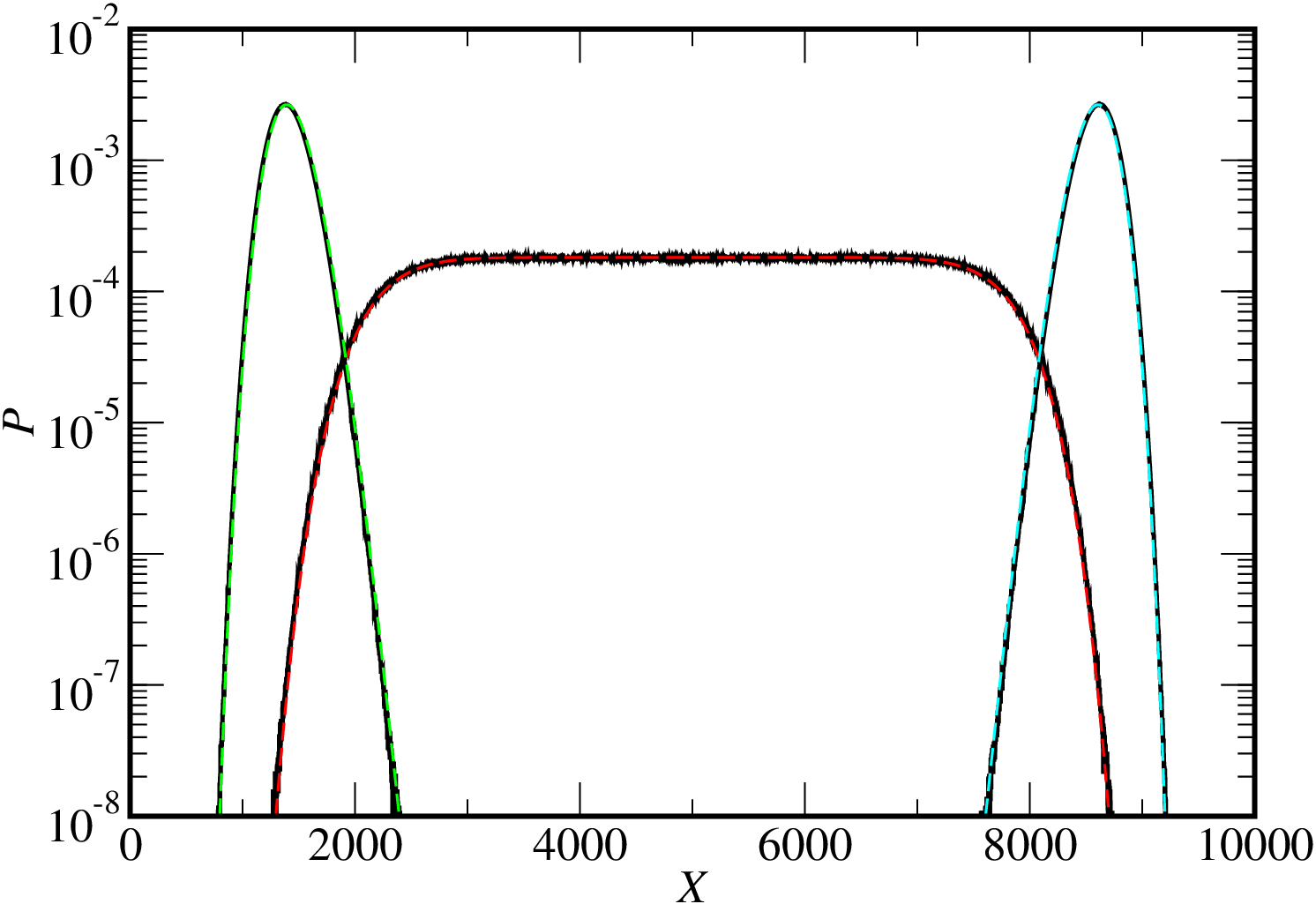}
\caption{Comparing XIE simulation with theoretical calculations: 
probability distribution $P$ of number of cross links $X$. Results 
are shown for $N_I=N_E=100$ (red),$N_I=101,N_E=99$ (green), and 
$N_I=99,N_E=101$ (blue). In each case, the dashed color line is 
the theory result and the thick black line enveloping it is the 
result of the corresponding numerical simulations.} 
\label{P(X)fig}
\end{figure}

Focusing on $f<0.5$ for now, this balance is spoiled by the presence of 
I's with no links, leading us to $\left[ 1-\rho _{I}\left( 0;X\right) \right] P\left( X\right) \cong P\left( X-1 \right) $ 
instead. In other words, as $X$ wanders towards $0$, the chances of 
it being `repelled' increases, hinting at the notion of the
`wall.' Of course, by symmetry, similar results can be obtained for the $f\simeq 1$ 
regime by finding a sum like Eqn. (\ref{Z(X)}) for $\rho_{E}\left( L;X\right) $, 
the probability that extrovert is fully connected.
Imposing the symmetric balance equation $\left[ 1-\rho _{I}\left( 0;X\right) %
\right] P\left( X\right) =\left[ 1-\rho _{E}\left( L;X-1 \right) \right]
P\left( X-1\right) $, we arrive at an analytic expression:%
\begin{equation}
P\left( X\right) \propto \prod \limits_{\Xi =1}^{X} \frac{1-\rho _{E}\left( L;\Xi-1 \right) } {1-\rho _{I}\left( 0;\Xi \right) }  \label{P(X)}
\end{equation}%

In Fig.\ref{P(X)fig}, we illustrate how remarkably well this prediction 
agrees with simulation data of the $L=100$ case. Further, it is
straightforward to generalize these considerations to the $N_{E}\neq N_{I}$
cases, e.g., by studying $\rho _{I}\left( k;X\right) \propto \left[ N_{I}\left(
1-f\right) \right] ^{-k}/\left( N_{E}-k\right) !$ In Fig. \ref{P(X)fig}, we
see that the results for $\Delta = \pm 2$ cases also agree spectacularly well with the data.

\begin{figure}[ht!]
\includegraphics[width=\linewidth]{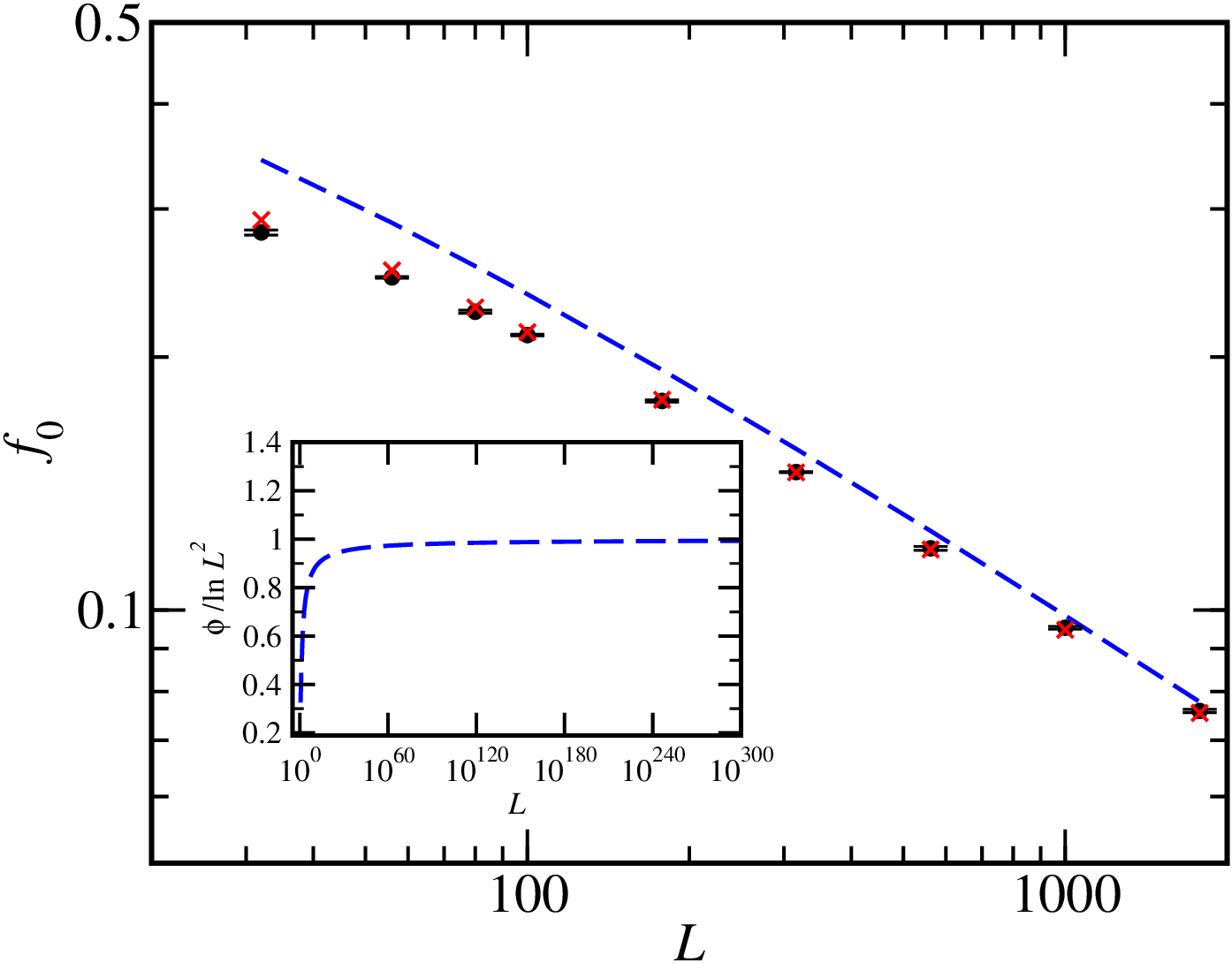}
\caption{Position of the (smaller) inflection point $f_0$ \textit{vs.} system size $L$. 
Black circles with error bars are simulation data. Red crosses are theoretical predictions. 
Blue dashed line is an approximate $f_0$ from solving a transcendental equation. 
Inset shows true asymptotics, setting in around $10^{100}$. }…
\label{X-wall}
\end{figure}

Exploiting this result for $P\left( X\right) $, we proceed to find the
position of the `wall' analytically. First, let us propose a natural place
to call `the edge of the mesa': the steepest decent as $P$ drops from the
`plateau' into the `plain.' 
These are also the inflection points of $P$: one near $f=0$ and the other, 
near $f=1$ as $L\rightarrow \infty $.
Denoting the former by $X_{0}$ (while the latter is just 
$L^{2}-X_{0}$ by symmetry), we see that it maximizes the gradient, $Q\left(
X\right) \equiv P\left( X\right) -P\left( X-1\right) $. For discrete $X$, 
$X_{0}$ can be defined as the value for which 
$\left\vert Q\left( X_{0}\right) -Q\left( X_{0}-1\right) \right\vert $ 
is smallest. Let us
approximate this by $Q\left( X_{0}\right) \cong Q\left(
X_{0}-1\right) $. Now, near $X=0$, we have $Q\left( X\right) =P\left( X\right) %
\left[ 1-P\left( X-1\right) /P\left( X\right) \right] 
\cong P\left( X\right) \rho_{I}\left( 0;X\right) $, 
so that the condition for $X_{0}$ reduces to a succinct one:%
\begin{equation}
Z\left( X_{0}\right) -Z\left( X_{0}-1\right) \cong 1  \label{Z}
\end{equation}%
In Fig. \ref{X-wall}, we see the excellent agreement between simulation data
(circles with error bars) and predictions from Eq. (\ref{Z}) (crosses). 
Of course, we notice the small discrepancies and ascribe them to the error inherent 
in our mean field approximation (replacing the stochastic $\ell$ by its average value $X/L$).

This approach also allows us to analyze the asymptotic behavior of $f_{0}=X_{0}/L^{2}$ as 
$L\rightarrow \infty $. It is clear that the largest terms in the sum 
in Eq.~(\ref{Z(X)}) occur around $\hat{q}=L\left( 1-f\right) $ and that the summand is
well approximated by a Gaussian: 
$\exp \left\{ -\left( q-\hat{q}\right) ^{2}/2\hat{q}\right\} $. 
Thus, the terms are effectively zero for 
$q-\hat{q}\gtrsim \sqrt{\hat{q}}\simeq \sqrt{L}$. Meanwhile, the sum extends beyond 
$\hat{q}$ by $Lf$. Thus, for $f>\sqrt{1/L}$ (which will turn out to be
satisfied), we can extend the sum to infinity and replace it by 
$\exp \left[ L\left( 1-f\right) \right] $. Further, in this limit, Eqn. (\ref{Z})
is just $L^{-2}\left. \partial _{f}Z\right\vert _{f_{0}}=1$ so that $\left[
e^{f_{0}}\left( 1-f_{0}\right) \right] ^{-L-1}e^{f_{0}}f_{0}=\sqrt{L/2\pi }$
. Assuming $f_{0}\ll 1$, dropping $O\left( 1/L\right) $ contributions, and
letting $f_{0}+\ln \left( 1-f_{0}\right) =-f_{0}^{2}/2+...$, we find a
transcendental equation for $\phi \equiv f_{0}^{2} L = X_{0}^{2}/L^{3}$%
\begin{equation}
\phi +\ln \phi =\ln L^{2}/2\pi   \label{trans}
\end{equation}%
Though it is tempting to conclude that, to leading order, 
$\phi \sim O\left( \ln L^2 \right) $ and 
$f_{0}\rightarrow \sqrt{\left( \ln L^{2}\right) /L}$,
such an estimate fails to fit the data for $L<2000$. For a similar reason,
we keep the $2\pi $, as $\ln 2\pi $ is comparable to $\ln 2000$. Instead,
when Eq.~(\ref{trans}) is solved numerically, the resultant $f_{0}$'s
appear to provide an increasingly tight upper bound to the data (dash line
in Fig.\ref{X-wall}). Our conclusions are clear: While the true asymptotics
of $f_{0}$ is $\sqrt{\left( \ln L^{2}\right) /L}$, this behavior does not
set in for the $L$'s we can access in simulations. Fortunately, for such 
$L$'s, Eq.~(\ref{trans}) provides reasonable bounds while Eq.~(\ref{Z})
agrees with data at the $1\%$ level for $L$ as small as $100$. To appreciate
how large $L$ must be before the true asymptotic form sets in, we show in
the inset of Fig.\ref{X-wall} a plot of $\phi /\ln L^{2}$, from the solution
of Eq.~(\ref{trans}), against $\ln L$, up to $L=10^{300}$. Even at $10^{100}$, 
this quantity misses unity by about 2\% ! Needless to say, we should not
expect to see simulations to confirm this asymptotic form in our lifetimes.

\begin{figure}[h!]
\includegraphics[width=\linewidth]{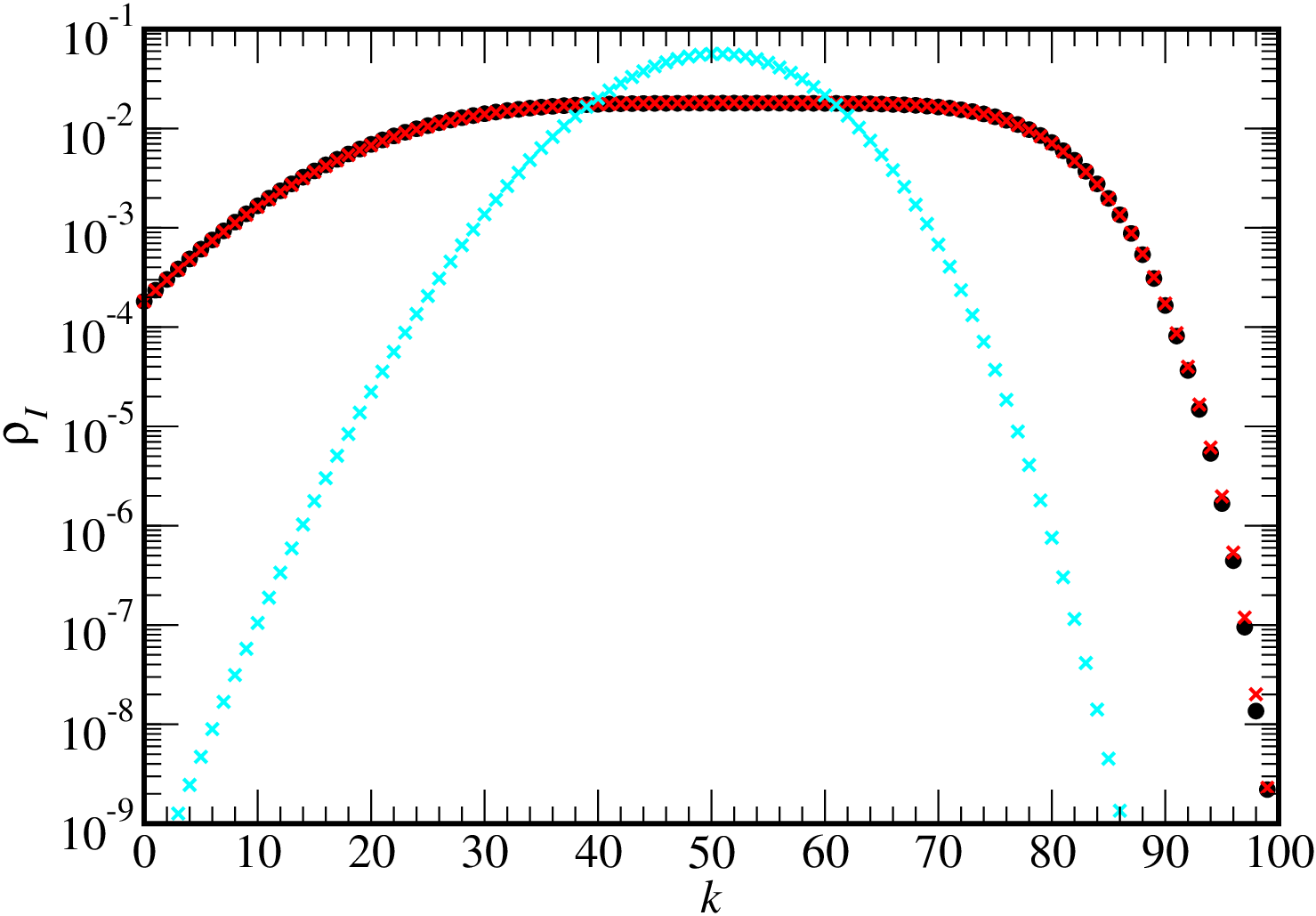}
\caption{Introvert edge degree distribution $\rho_I(k)$ for the $N_I=N_E=100$ case. The Black circles are the simulation data. Red crosses are predictions from the present theory. Blue crosses are results from an earlier theory (SCMF). } 
\label{criticalDD}
\end{figure}

We end this section with the resolution of another issue in the $XIE$ model:
the disagreement between the SCMF prediction and data in the degree
distributions of the critical $L=100$ case (e.g., blue crosses and black circles 
for $\rho _{I}\left( k\right)$ in Fig. \ref{criticalDD}). 
Let us focus on an I again and note that 
$\rho _{I}\left( k\right) =\sum_{X}\rho _{I}\left( k;X\right) P\left( X\right) $. 
With expressions (\ref{rho(k;X)}) and (\ref{P(X)}), we have a new
prediction for $\rho _{I}\left( k\right) $. Though somewhat cumbersome, it
is simple to carry out the sum numerically. 
Plotted as red crosses in Fig. \ref{criticalDD}, we again find remarkably excellent agreement with data.

\section{Conclusions and Outlook}

Since its discovery \cite{LSZepl2012}, the extraordinary variability in $X$, the
number of links between an equal number of extreme introverts and
extroverts, has remained a theoretical puzzle. In this letter, we presented
a new perspective and an associated approximation scheme which proved
successful in solving this puzzle. Unlike earlier approaches, we 
considered ensembles with fixed $X$, much like Ising models with conserved magnetization. 
We are motivated to take this approach by two observations. One is the success 
of the SCMF theory \cite{BLSZpre15} for all but the critical system. 
The other is that correlations between the microscopic variables $a_{ij}$ appear to be minimal~\cite{EZBZ18}. 
Thus, the conjecture is that, despite the presence of long-range and multi-spin interactions in $\mathcal{H}$, 
the large variations in $X$ for the critical system are \textit{not} in conflict with the applicability of a mean field treatment. 
Such a conjecture leads us to to approximate the stochastic $\ell$ with its average $X/L$ and to the subsequent successes. 
Further along these lines, we believe that any observable quantity $\mathcal{O}$ 
(which has a limited variability in a fixed $X$ ensemble) will display an extreme Thouless effect.
The reasoning is that its statistics will be `carried' by $P(X)$, so that its average will suffer a maximal discontinuity across criticality while it variability will also be maximal at criticality. 

\begin{figure}[h!]
\includegraphics[width=\linewidth]{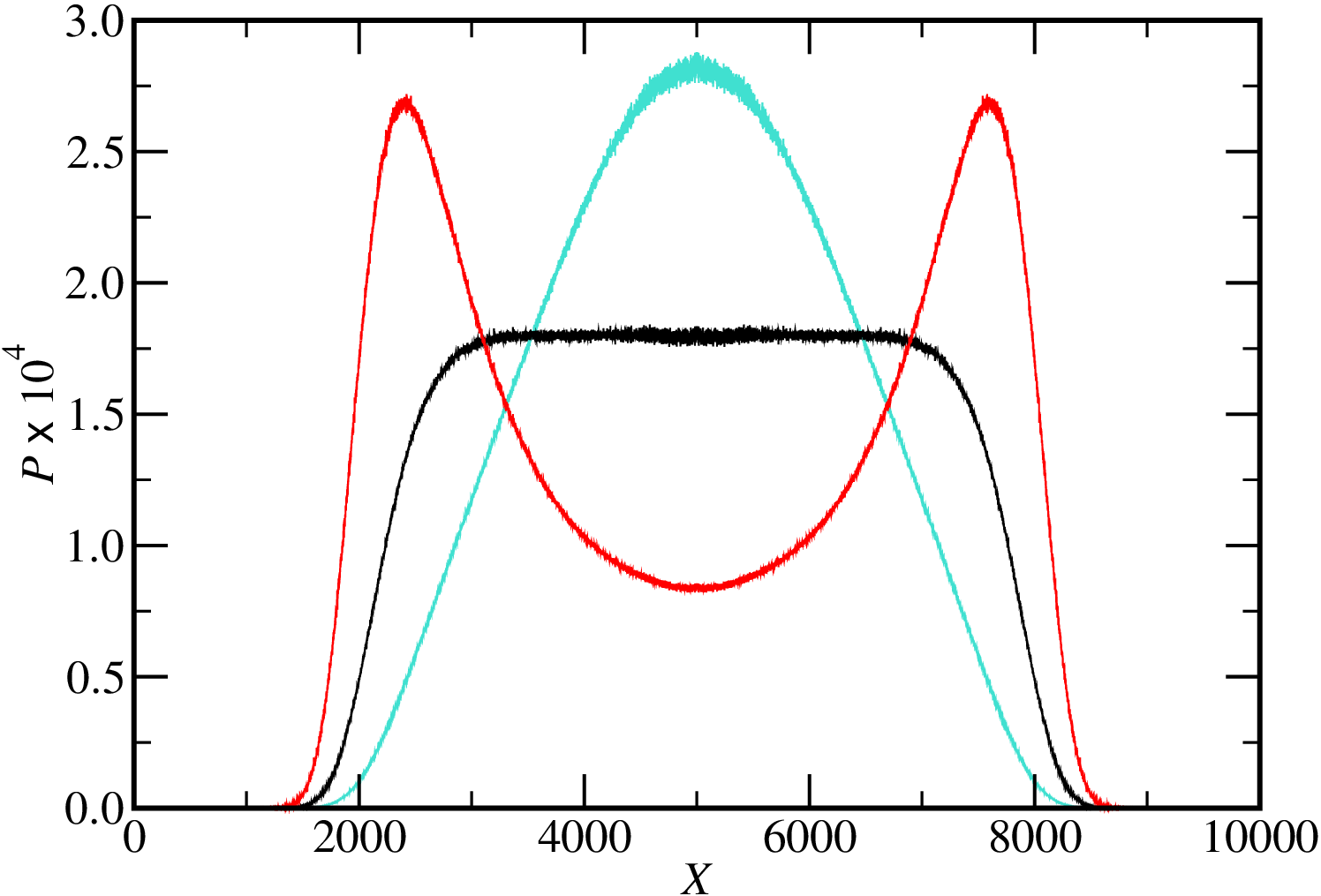}
\caption{Probability distribution $P$ of number of cross links $X$ for different values of $\beta$: Black shows simulation results for the critical value $\beta=1$, blue the results above criticality at $\beta=0.99$, and red the results below criticality at $\beta=1.01$. The off-critical results were obtained from the simulation at criticality using the method of Ferrenberg and Swendsen~\cite{Ferrenberg88}.} 
\label{varbeta}
\end{figure}

These considerations dispel another `rule of thumb' in phase transitions: 
an intimate connection between the large fluctuations of the OP and sizable correlations among the microscopic variables. 
Here, we see that this link is severed in $XIE$, if only through some `conspiring' interactions in $\mathcal{H}$. 
In this spirit, we believe a simple lesson can be learned by considering the following. 
If we start with a \textit{non-interacting} Ising model, then the distribution of the total magnetization 
$M = \sum s_i$ is just the binomial: $P(M) \propto \binom{N}{N_{+}}$, with 
$N_{\pm} \equiv (N \pm M)/2$.
If we now impose a `Hamiltonian' of the form
$\mathcal{H}_M = -\ln N_{+} ! - \ln N_{-}!$,
then the resultant $P(M)$ is completely \textit{flat}, so that the 
variability in $M$ is maximal~\footnote{This system is precisely 
the one studied in ref. \cite{FFK16}, arrived at from a minimal model of spin dynamics.
Unlike $XIE$, it is trivially solvable, since it effectively reduces to 
a statistical mechanical system with a single variable, $M$.}. 
Nevertheless, regardless of the apparent existence of 
`long-range and multi-spin' interactions in $\mathcal{H}_M$, correlations 
between the spins are `trivial' (e.g.,  $\left\langle s_{ij} \right\rangle $ 
being just 1/3 for all $i \neq j$, 
in contrast to power law decays in the critical region of the ordinary Ising model).   
Moreover, we can play the game of statistical mechanics further, by adding 
temperature and magnetic field to a Boltzmann factor:
$ exp{- \beta [\mathcal{H}_M - HM]} $.
Then, we can expect an extreme Thouless effect at the `critical point' ($\beta=1,H=0$). 
The same can be done for the $XIE$ model by multiplying a temperature-like parameter, 
$\beta$, to that `Hamiltonian', $\mathcal{H}$.  
Illustrated in Fig.~\ref{varbeta}, preliminary results show that 
$P(X)$ displays the expected features: 
single-peaked for $\beta =0.99$ (`above criticality')
and bimodal for $\beta =1.01$ (`below criticality'). 
Work is in progress to explore these ideas in a systematic way, as well as more 
realistic social networks (e.g., with generic numbers for preferred contacts 
rather than $0$ and $\infty$). We believe these studies are valuable not only 
for further understanding of the XIE model, but also for providing insight 
into the Thouless effect (extreme or more generic) in other systems,
as well as painting a more complete picture of the subtle 
characteristics of phase transitions in general.

\begin{acknowledgments}
We thank D. Dhar and B. Schmittmann for illuminating discussions, and F. Greil for 
his efforts during the initial phases of this project. This research is supported
by the US National Science Foundation, through grant DMR-1507371. 
\end{acknowledgments}

\bibliography{WallPaper,PRE-RZ1}








\end{document}